\documentclass[12pt,preprint]{aastex}

\shorttitle{}
\shortauthors{}

\begin{document}

\title{Quiet Sun X-rays as Signature for New Particles}

\author{K. Zioutas\altaffilmark{1,4,5},
        K. Dennerl\altaffilmark{3},
        L. DiLella\altaffilmark{4}}
\author{D.\,H.\,H. Hoffmann\altaffilmark{1},
        J. Jacoby\altaffilmark{2},
        Th. Papaevangelou\altaffilmark{1}}

\altaffiltext{1}{Institut f\"ur Kernphysik, TU--Darmstadt,
                 Schlo{\ss}gartenstr.~9, D--64289 Darmstadt, Germany;\\
                 \phantom{xxxx}{\tt hoffmann@physik.tu-darmstadt.de,
                 Thomas.Papaevangelou@cern.ch}}
\altaffiltext{2}{Institut f\"ur Angewandte Physik, Johann Wolfgang
                 Goethe Univ., D--60486 Frankfurt am Main, Germany;\\
                 \phantom{xxxx}{\tt Jacoby@Physik.Uni-Frankfurt.de}}
\altaffiltext{3}{Max--Planck--Institut f\"ur extraterrestrische Physik,
                 Giessenbachstra{\ss}e, D--85748 Garching, Germany;\\
                 \phantom{xxxx}{\tt kod@xray.mpe.mpg.de}}
\altaffiltext{4}{CERN, CH--1211 Geneva 23, Switzerland;\\
                 \phantom{xxxx}{\tt Luigi.Di.Lella@cern.ch}}
\altaffiltext{5}{Physics Department, University of Thessaloniki,
                 GR--521 14 Thessaloniki, Greece;\\
                 \phantom{xxxx}{\tt zioutas@cern.ch}}

\begin{abstract}
We have studied published data from the {\it Yohkoh} solar X--ray mission, with
the purpose of searching for signals from radiative decays of new, as yet
undiscovered massive neutral particles. This search is based on the prediction
that solar axions of the Kaluza--Klein type should result in the emission of
X--rays from the Sun direction beyond the limb with a characteristic radial
distribution. These X--rays should be observed more easily during periods of
quiet Sun. An additional signature is the observed emission of hard X--rays by
SMM, NEAR and RHESSI. The recent observation made by RHESSI of a continuous
emission from the non--flaring Sun of X--rays in the 3 to $\sim 15$~keV range
fits the generic axion scenario. This work also suggests new analyses of 
existing data, in order to exclude instrumental effects; it provides the
rationale for targeted observations with present and upcoming (solar) X--ray
telescopes, which can provide the final answer on the nature of the signals
considered here. Such measurements become more promising during the
forthcoming solar cycle minimum with an increased number of quiet Sun periods.
\end{abstract}


\section{Introduction}

In order to solve the strong CP problem (why the strong interaction contrary
to weak interactions does not violate CP symmetry), a new neutral particle,
the axion, with spin--parity $0^-$ has been invented \citep[for recent review
articles see, e.g.][]{raffeltg99,raffeltg00,bradley}. Axions, along with
Weakly Interacting Massive Particles (WIMPs), are the two leading particle
candidates for dark matter in the Universe. If axions exist, they should be
abundantly produced inside the solar core. An axion can be seen as a very
light neutral pion ($\pi ^0$), with restmass  $\sim 10^{-3\pm 3}$ eV/c$^2$.
However, while the  $\pi ^0$ interacts strongly and decays to two photons with
a mean lifetime of $\sim 10^{-16}$ s, the axion interacts very feebly with
matter and is expected to decay to two photons ($a\rightarrow 2\gamma$) with a
lifetime much longer than the age of the Universe. Therefore, detection
techniques utilise the axion interaction with the electric field of atoms of
underground detectors or strong magnetic fields, in order to (coherently)
convert them to real photons. In other words, electric or magnetic fields play
the role of a catalyser, which can  transform axions into detectable photons.
If very light axions ($m \leq $ few eV/$c^2$) are produced inside the Sun,
their thermal energy peaks at $\sim 4.2$ keV and they are ultra-relativistic.
They mostly stream out of the Sun and can have an impact only on the evolution
of a Star through the additional escaping energy (as it happens with
neutrinos).
        
However, in recent theories of extra--dimensions, proposed as extensions of the
Standard Model, the `conventional', almost massless axions become as massive
as the reaction energies involved. In the case of the solar axions, the
expected mass spectrum of all the excited Kaluza--Klein (KK) states extends all
the way to $\sim 10$ keV/c$^2$ \citep{kk,dl,dilella}. These high KK--masses
imply a relatively shorter lifetime ($\tau \sim 10^{20}$s), because of the
$\tau \sim m^{-3}$ dependence. The underlying axion--photon--photon coupling
constant, $g_{a\gamma \gamma}$, which defines the interaction cross-section
with ordinary matter, is the same for the `conventional' almost massless axion
and for the massive KK--axions. In this work, the KK--axions are taken as a
generic example of particles which can be created inside the hot solar core. A
small fraction of them ($\sim 10^{-7}$, as it has been estimated by
\citet[hereafter DZ03]{dilella}), are extremely nonrelativistic and they can
be gravitationally trapped by the Sun itself in orbits where they accumulate
over cosmic times. As shown in DZ03, their density increases enormously near
the solar surface. The estimated mean distance of the KK--axion population from
the Sun surface is $\sim 6.2\,R_{\odot}$.

If axions of the KK--type are gravitationally trapped and decay to two photons,%
\footnote{Following the decay mode $a\rightarrow \gamma \gamma$, the two
photons have the same energy and are emitted back--to--back, because of
momentum conservation, since they are highly non-relativistic.}
then the observed X--ray surface brightness from the solar disk and limb can be
a signature of the solar axion scenario. In this work, we discuss the expected
surface density profile of the derived axion halo around the Sun (DZ03), and
we compare it with published data  taken by the {\it Yohkoh} soft X--ray
telescope from the diffuse emission of two quiet Sun observations. In
addition, the measured continuous emission of hard X--rays (below
$\sim15\mbox{ keV}$) from the non--flaring Sun during the past $\sim25$ years
by different orbiting X--ray detectors is also considered.

\section{The Signatures}

\subsection{Radial X--ray distribution}

Assuming a spatial density distribution of radiatively decaying particles in
the solar periphery as described in DZ03, we calculate the surface X--ray flux
expected from inside and outside the solar disk by integrating over the line
of sight (see Figure 1). In doing so, if the observed X--rays from the Sun
direction are exclusively due to the radiative decay of massive
non--relativistic particles along the pointing direction, their density
distribution can thus be derived. This is actually a standard procedure
applied in Astrophysics. The place around the solar limb, i.e.\
$R=1\,R_{\odot}$, provides a characteristic change in surface luminosity,
because of the rapid change of the integration path length of the strongly
absorbed photons from axion decay.

Figure 2 shows again the axion simulation used in Figure 1, covering an
elongation angle up to  $\sim 50^{\circ}$ away from the Sun. The level of the
cosmic X--ray background (XRB) relative to that expected to come from the Sun
is also shown. The XRB defines in fact the size of the predicted solar X--ray
halo within the axion model; for the hard and soft energy band, it extends
`only' to $\sim 3.5^{\circ}$ and $\sim 9^{\circ}$, respectively.

We suggest, therefore, that the detection of an extended X--ray halo around the
Sun is a signature of KK--axions produced in the core of the Sun, if the
contributions from conventional solar dynamical effects, e.g.\ flares, can be
excluded. The X--ray surface brightness resulting from the decay of trapped
axions is expected to be continuous and stable with time and to decrease
exponentially with increasing distance from the Sun (see Figure 1 and 2). At
first, this signature can be studied better during quiet Sun conditions,
because any contributions from the dynamic Sun can thus be minimized. These
conditions correspond to the {\it Yohkoh} solar data which we have used in
this work. However, beyond a certain distance from the solar limb where even
the impact of the active Sun becomes negligible, one could study continuously
the X--ray emission from the periphery of the Sun. The next solar cycle minimum
with extended quiet Sun periods can be utilized to perform such measurements.
The field--of--view (FOV) of the X--ray telescope should cover the largest
possible distances from the solar disk. The limited FOV of all imaging solar
X--ray observatories ($\sim 1^{\circ}$ at best) can be effectively widened by
pointing away from the Sun centre.

Figure 1 shows the spatial distribution of two observations made by the {\it
Yohkoh} imaging detector in the soft X--ray band ($0.25 \leq E_{\gamma} \leq 4$
keV) \citep[hereafter SWA96 and WSA97]{region1,region12}. To the best of our
knowledge, these are radially the most extended measurements of this kind from
quiet Sun regions published so far. These two regions were assigned as being
`quiet' before this work, i.e.\ the selection was unbiased, and, it appears
that the approach of SWA96 and WSA97 was reasonable. However, it is crucial
that possible contributions from active regions on the solar disk into the
limb region due to scattering in the X--ray telescope be excluded, because
they give rise to a radial distribution similar to that expected from the
decaying axions in the solar halo. The treatment of such far--scattering
effects, based upon a correction made by \citet{hara} on the {\it Yohkoh} PSF,
as it was determined by \citet{martens}, reached the conclusion that bright
nearby regions constitute a small but not insignificant fraction (WSA97).

The bulk of the published work from all solar instruments, operating since
years in space, focuses onto the flaring Sun, with the quiet Sun being almost
overlooked. However, the soft X--ray energy band of the {\it Yohkoh}
observatory overlaps with the lower energy range of the axion model, which
gives a broad distribution peaking at $\sim 5 $ keV (DZ03). Then, these soft
X--rays can be associated with particles of rest mass 2$\times E_{\gamma}$  (a
few keV), i.e.\ they {\it may} be used to test the axion scenario (DZ03),
extending it thus to lower energies, even though we can not argue
quantitatively for this. From the solar axion Monte Carlo simulation, the
radial distribution of the assumed KK--axions is actually independent on the
axion restmass. Therefore, in this work we look first at the shape of the
radial distribution of measured solar X--rays and we compare it with the shape
predicted by the axion model. Of course, once X--ray observations at higher
energies become available, they will allow a more quantitative comparison
between prediction and observation.

The region chosen during the May 1992 observation was cleaned from so--called
X--ray brightenings, which were three small regions with enhanced X--ray
intensity. Keeping in mind the off--limb scattering considerations given above,
this cleaning might explain the surprisingly high degree of agreement of the
shape of the radial distribution of the May 1992 run with that expected by the
axion model, in comparison with the August 1992 observation (see Figure 1).
Therefore, the May 1992 measurement could be taken as the kind of evidence of
the axion  scenario we search for, if the flux diffusing out beyond the limb
is really negligible. Therefore, the off--limb scattering issue must be looked
at much more closely.  
	In fact, Fig. 2 of the work by Sturrock et al. 1996 provides the estimated 	scattering component also across the solar edge, which makes only a 
      $\sim$ 25$\%$ change. The tail(s) of the PSF from one or more bright 	points on the solar disk could in principle result to the observed X-ray 
      intensity off the limb. However, it seems actually difficult to 
      reconcile this picture with the PSF tail(s) with the measured change in 	brightness by as much as by a factor of 4 across the solar disk. 	Therefore, compared to the August 1992 run, the evaluated data from May 	1992 are distinguished for the reasoning of this work.
The measurement from August 1992 has the advantage of
increasing the radial distance by a factor of 2, thus covering one solar
radius beyond the limb. However, the agreement with the shape of the axion
model is certainly not as good as it is the case for the May 1992 run. This is
not surprising, since such observations are affected by uncertainties such as
the level of contamination by hidden active regions (see above), or
contributions from Bremsstrahlung and/or synchrotron radiation by the corona
plasma, etc.

In addition, these observations were performed during a solar maximum, and the
quiet Sun is not as quiet as during a solar minimum. Therefore, the observed
degree of agreement between the two {\it Yohkoh} observations and the axion
model is even more surprising. To put it differently, let us assume that the
variations between the two experimental curves in Figure 1 reflect the
underlying fluctuations (whatever the reason) between two independent
measurements. Although only these two  measurements are available, it is not
unreasonable to assume that the underlying distribution should be some curve
between them. Then, it would be more reasonable, because less biased, to
compare our predictions with a curve obtained by the average of the two {\it
Yohkoh} observations. In fact, these two observations made $\sim 3.5$ months
apart show fluctuations from the average of some $\pm 30 \% $, while the
radial X-ray brightness changes by as much as a factor of $\sim 10$.

\subsection{Quiet Sun hard X-ray emission}

Interestingly, it has been recognized that the solar outer atmosphere is
somehow continually and globally heated to MK temperatures, i.e.\ this heating
is a ubiquitous quasi steady--state activity \citep{moore}. We discuss below
quiet Sun X--ray observations with photon detectors in space. To find such data
is not an easy task, since all relevant missions are primarily interested in
solar flares. However, for this work such observations provide important
information on the radial X--ray distribution.

The emission of X--rays above 3.5~keV from the non-flaring Sun was first
observed in the 1980's by the HXIS spectrometer on SMM \citep{simnet,simnet1}.
Later (1996--2001), the NEAR spacecraft mission to the asteroid 433 Eros
provided quiescent solar X--ray spectra: a) directly with two solar monitors
\citep{near1}, and b) indirectly from the scattered solar X--rays by Eros
\citep{near2,starr}. The analog X--ray spectra cover the $\sim 2$ to $\sim 10$
keV range.

The RHESSI observatory \citep{hurford,lin} has confirmed these results: it has
observed a {\it continuous} X--ray emission, from 3 to $\sim15$~keV, with
frequent (every few minutes) microflaring \citep{rhessix1a}, when there are no
observable flares present \citep{rhessix2a}. This continuous emission of hard
X--rays is consistent with the solar axion scenario behind this work. For
example, Figure 3 gives the measured X--rays by the GOES and RHESSI missions
in orbit: a jump in the low energy count rate between 3 and 12~keV when the
RHESSI day starts or ends, is clearly visible \citep{rhessix3}. The GOES X--ray
satellite intensity above 3.1~keV is also at its lowest level (around 02:20h),
and there is no sign of a flaring Sun from both orbiting instruments. Since
during this period the RHESSI counting rate is clearly above the detector
background level, this excess X--ray intensity comes from the non--flaring Sun.

Additional observations can be found in \cite{rhessix1b}. Moreover, a
statistical observation has been made\,%
\footnote{In the period from October 2002 through May 2003.}
with $\sim 1000$ such measurements \citep{rhessix2b}. The measuring time--bin
is one minute\,%
\footnote{For comparison, the time resolution of the detector is $\sim10$ msec.}
during detector daytime (with the Sun being in the RHESSI FOV), and its low
energy counting rate is taken at a minimum. In order to avoid microflares, a
statistical flatness check was applied one minute before and after each
interval. For the so defined ``quiet Sun''  a temperature distribution of
$\sim 6$\,--\,8~MK is derived. Such temperature values go well above the
usually quoted quiet Sun corona temperature of $\sim$ 1 to 2~MK, and they may
require careful interpretation.%
\footnote{Note that the applied procedure is valid within the concept of
nanoflares \citep{rhessix4} (i.e.\ frequent small energy bursts), which is not
strictly compatible with the observed continuous hard X--ray emission
\citep[for a critical consideration of the advocated nanoflare hypothesis,
which {\it `seems to be close to be falsified', see}][]{berghmans}, predicted
this work. For example, it has been already noticed that the resulting
Differential Emission Measure does not show a predicted peak near 3~MK
\citep{rhessix2b}. This might not be a surprise, since in related calculations
``the effective line-of-sight thickness of the coronal plasma (or height for
observations in the center of the disk) cannot be measured and must be
assumed'' \citep{mitrabenz}, while the presence of possible plasma
nonuniformities can further complicate the conventional interpretation
\citep{rhessix4}.}

\subsection{Numerical estimates}

From Figure 3 one can perform an approximate intercalibration between RHESSI
and GOES, by using, for example, flare $\# 2$. The minimum rate above detector
background around 02:15h is $\sim 50$ cts/s, highly significant statistically.
This corresponds to a brightness of $\sim 2.6\cdot 10^{-10}$Watts/m$^2$; i.e.,
the  non--flaring solar X--ray luminosity in the range 3\,--\,12~keV is $\sim
6\cdot 10^{20}$~erg/s. One should note that the corresponding value used in
the solar axion model was $\sim 10^{23}$~erg/s in the energy range
2\,--\,8~keV. Having in mind the relation\,%
\footnote{Combining relation (1) and (2) in DX03.}
between the X--ray Luminosity and the axion--to--photon coupling constant $L_x
\sim (g_{a\gamma \gamma})^4$, the lower X--ray luminosity derived above
implies a smaller coupling constant: $g_{a\gamma \gamma} \geq 2.5\cdot 10^{-14}
\mbox{GeV}^{-1}$. However, we made the assumption that flare $\# 2$ has the same
X--ray energy spectrum as that from the non--flaring Sun. In fact, flares emit
higher energy X--rays (as can be seen also in Figure 3 (bottom)), while the
RHESSI detection efficiency increases very rapidly with energy. This tends to
underestimate the photon intensity for the non--flaring Sun, and therefore, the
above value for the coupling constant ($g_{a\gamma \gamma}$) is a safe lower
limit (a value of $g_{a\gamma \gamma} \sim $ 4-5$\cdot 10^{-14}$GeV$^{-1}$
might be reasonable).

Similarly, in order to derive a limit of the coupling constant $g_{a\gamma
\gamma}$ from the {\it Yohkoh} observations, we assume that the axion model
can be extended to the lower energy band of {\it Yohkoh} ($E_{\gamma} \geq
0.5$ keV). The estimated quiet Sun X-ray luminosity between $\sim 0.5$ and
$\sim 4$~keV is\,%
\footnote{Note that $\mbox{1 DN} = 5.8\cdot10^{-10}\mbox{ erg}$ at the SXT
focal plane and that the pixel size is 2.5 arcsec \citep{tsuneta,yohkohz},
i.e.\ the Sun disk makes $\sim 4\cdot 10^5$ pixels. Furthermore, the following
input parameters were used in order to estimate the whole Sun soft X--ray
luminosity \citep[SWA96, WSA97]{tsuneta,yohkohz}: intensity = 1200 DN/pixel,
SXT effective area $\approx 0.4$ cm$^2$, energy band width $\approx 2.3$ keV
and exposure time = 136.5~s. The luminosity so obtained has been normalised to
the same energy band width used at higher energies in the axion model (DZ03).
For comparison, if we assume that 1 DN corresponds to $1.5\cdot 10^{19}$ erg at
the Sun \citep{tsuneta,yohkohz}, we arrive to a similar luminosity (however,
under the assumption of a $\sim 3$MK plasma). Note that RHESSI measured quiet
Sun temperatures $\sim 3$ times higher \citep{rhessix2b}, which should result
in a lower luminosity. Therefore, the applied procedure is, so to say, model
independent.}
$L_x\approx 3.75\cdot 10^{25}$~erg/s. This much higher solar soft X--ray
brightness implies, again according to the $L_x \sim (g_{a\gamma \gamma})^4$
relation, a larger coupling constant, i.e.\ $g_{a\gamma \gamma} \leq 40\cdot
10^{-14}\mbox{ GeV}^{-1}$, in order to achieve a normalization between the two {\it
Yohkoh} observations and the axion model (DZ03) shown in Figure 1. The value
of $g_{a\gamma \gamma}$  derived from the $\sim 1$ keV range is still $\sim
500$ times smaller than the best experimental limit obtained by CAST
\citep{dieter}. For comparison, the corresponding value used in the solar axion
model is $g_{a\gamma \gamma} = 9.2\cdot 10^{-14}\mbox{ GeV}^{-1}$. The implied
simplifications or/and assumptions with the RHESSI and {\it Yohkoh} results
provide a wide range of allowed values for the coupling constant:

$2.5\cdot 10^{-14}\mbox{ GeV}^{-1} ~\leq ~g_{a\gamma \gamma}~ \leq ~40\cdot 10^{-14}
\mbox{ GeV}^{-1}$

\noindent with the axion model value being within these loose bounds.  Notice
that these are rough numerical estimates, and  the solar axion model is a
generic one.

\section{Discussion}

The discovery by \citet{grotrian} that the solar corona is $\sim 100\times $
hotter than the underlying photosphere has been a puzzle since then.
Remarkably, recent experimental work in this field arrives to the conclusion
that  {\it `\dots\ the results have made the coronal heating process even more
of a mystery'} \citep{antiochos}. Various conventional approaches \citep[see,
e.g.,][]{suzuki} to understand this strange finding have been suggested in
order to explain the coronal heating. The physical origin of the coronal
heating remains one of the most fundamental problems in stellar (and solar)
astrophysics \citep{guedel}.

The solar axion scenario  provides a continuous and steady energy input into
the solar atmosphere. However, depending on the local physical conditions,
e.g.\ magnetic field strength, plasma density, etc., an additional
axion--to--photon conversion mechanism may enhance locally this energy input.
Also, the derived density of the axion cloud is highest near the surface of
the Sun. It is conceivable that  with such a high density of boson states some
additional as yet unforeseen effects might occur, which may be influenced by
the physical environment. Then, the phenomenology associated with more
elaborated  models might provide important feedback to the axion scenario. We
mention, for example, models of small--scale reconnection processes
\citep[e.g.][]{sturrock}, which have seen much discussion in recent years
\citep{suzuki}. These models might point to a specific local property of the
solar atmosphere, which can have an impact on the ubiquitous trapped axions.
Having in mind the solar neutrino problem, it might well be that the solar
corona problem points also to a solution based on non--standard (solar)
physics. It is this approach that we follow in this work.

The radial distribution of the two independently measured {\it Yohkoh} images
in soft X--rays and the axion model predictions (Figure 1) appear quite
similar, suggesting a possible correlation between them. Interestingly, the
findings of the same two quiet Sun investigations (SWA96, WSA97) also favour a
mechanism that deposits {\it somehow} nonthermal energy as heat beyond the
observed range of heights above the limb (R$\geq 1.5$-2$\,R_{\odot}$) in the
quiet corona, consistent with an inward heat flux. Within the adopted model
(SWA96, WSA97), there is no evidence for appreciable energy input over this
radial range. However, an axion scenario, like the one used in this work, is
consistent with these observations, including the continuous emission of hard
X--rays from the quiet solar atmosphere. The observed X-rays below $\sim
15$~keV are much more energetic than the bulk of thermal photons from a $\sim
2$~MK solar corona plasma (e.g.\ $E_{\gamma}\approx 3kT\approx 0.5$ keV).

\section{Conclusion}

Our work suggests that the importance of quiet--Sun solar X--rays should not be
overlooked. This paper should provide motivation to the solar X--ray community
to follow experimentally the arguments presented here and to reduce the
uncertainties of relevant observations. The period during the (next) solar
cycle minimum should provide still cleaner quiet Sun  conditions. For example,
 offpointing observations with the RHESSI solar X-ray telescope can result to
interesting data above $\sim 3$ keV.~Even the pointing tests of the spacecraft
on the 8 and 23 May 2003 might well contain useful solar data, in spite of the
announcement that `data in this period should not be used for solar studies'\,%
\footnote{http://hesperia.gsfc.nasa.gov/hessi/ and
 http://hessi.ssl.berkeley.edu/$\sim$dsmith/hessi/HME.html}
$\!\!$. Similarly, we suggest that the announced offpointing to the Crab Nebula from
16 June 2003 should be evaluated in the light of this work. From Figure 2 and
the energy threshold of the RHESSI detectors, it follows that the relevant
angular extension from the surface of the Sun is approximately $\theta \leq
2^{\circ}$.

Other non--solar X--ray telescopes can be of potential use for this work,
provided they can reduce the elongation angle with the Sun. Then, instead of
searching for solar axions only here at the site of the Earth, (solar) X--ray
telescopes could provide new physics results associated with the solar X--ray
halo. At present, these orbiting instruments are highly sensitive to X--rays
below $\sim 10$ keV, which is  within the region-of-interest for many
astrophysical objects.

In order to explain the experimental approach of this work in a model
independent way, an exotic effect like the axion scenario of this paper can be
searched for in the following type of solar observations, provided
conventional solar dynamical effects can be excluded: a) an off--pointing
observation results to an increased radiation level while approaching the Sun.
This should be the manifestation of a halo around the Sun {\it whatever} its
radial distribution, and/or b) the measured level of {\it any} cosmic
radiation from the blank sky does not diminish completely (according to the
performance of the telescope) while pointing towards the solar disk, i.e.\
solar shadow effect is missing. The search for such a residual radiation from
the solar disk direction might be the most sensitive approach, especially at
photon energies above a few keV, where practically no emission is expected from
the quiet Sun.

{\it In summary}, the radial distributions from the re--considered two {\it
Yohkoh} X--ray  observations of the quiet Sun (including the derived inward
heat flux in the solar atmosphere of {\it some} nonthermal energy deposition
beyond $\sim 1$ solar radius from the Sun surface) can be reconciled with a
halo of decaying massive particles near the solar surface. New analyses of
existing data may definitely clarify that instrumental scattering effects from
bright points on the solar disk are  small. More extended radial X--ray
distributions (preferentially during a solar cycle minimum) can provide
important information for the quiet Sun. An additional and independent
evidence in favour of the axion scenario is the observed {\it continuous}
emission of X-rays from the non-flaring Sun in the 3 to $\sim 15$ keV range.
All these observations, when considered together, might suggest a
non--conventional mechanism of the type assumed here. New X--ray measurements
could clarify the nature of these relatively high energy effects around the
Sun, for which an alternative conventional explanation is still missing.

\acknowledgments

We thank the anonymous referee for his constructive and extensive remarks,
which helped us to further elaborate our work. The critical remarks by Jim
Rich are greatfully acknowledged. We also thank George Simnett for informative
discussions, and in particular, for allowing us to use results from
unpublished data. Our thanks go to Tullio Basaglia from the CERN central
library, who helped us enormously in finding  within short time various not
easily accessible publications. KZ, DHHH, JJ, ThP like to acknowledge the
support of BMBF/Germany under grant number 05CC1RD1/0.

\newpage

\vskip3.0cm
\begin{figure}[H]
\centerline{\includegraphics[width=20cm]{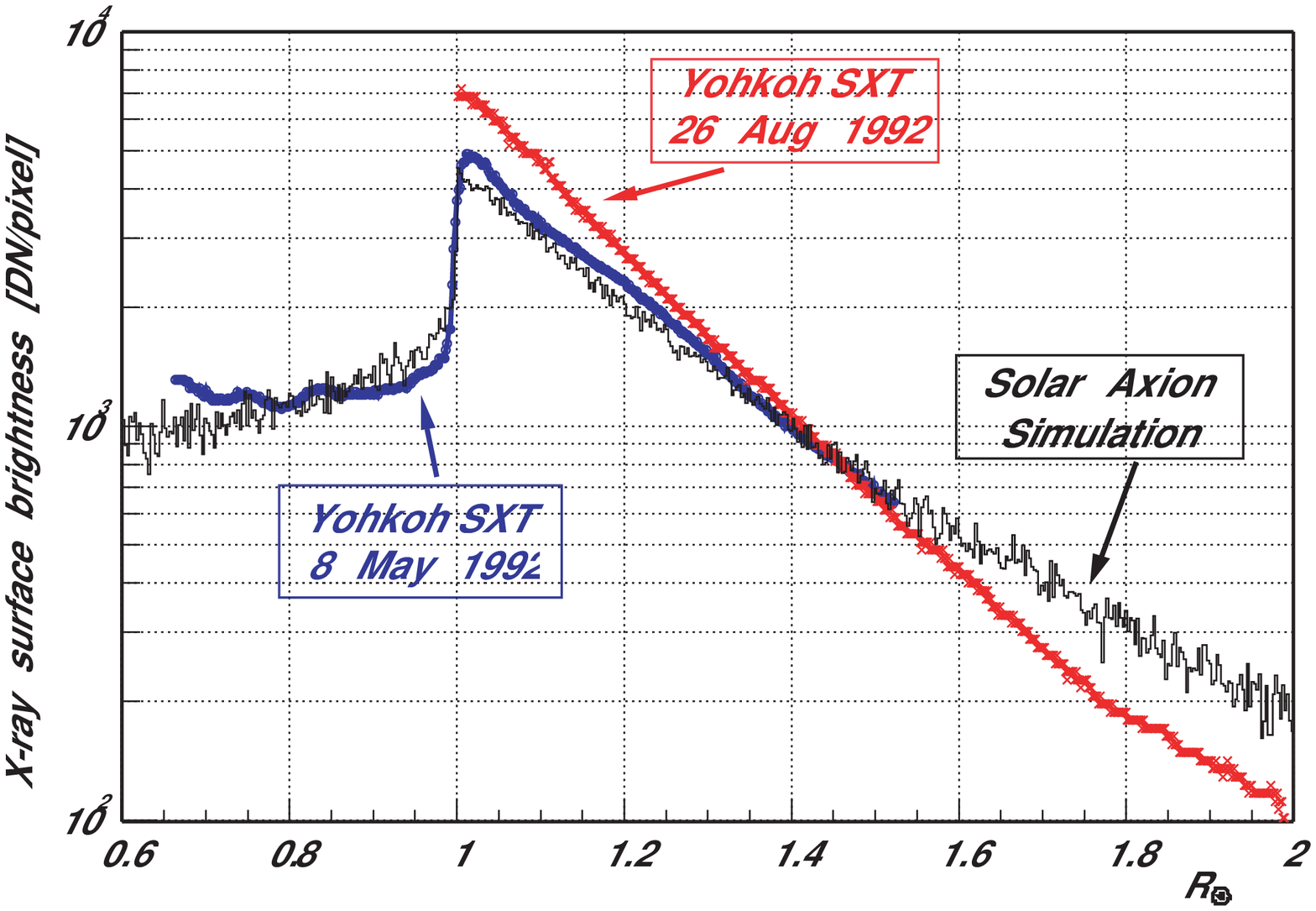}}
\caption[]{Theoretical (DZ03) and experimental \citep{region1,region12} soft
X--ray surface flux distributions from the quiet Sun. The simulated curve has
been shifted relative to the experimental points of both observations, which
implies $g_{a\gamma \gamma} \leq 40\cdot 10^{-14}\mbox{ GeV}^{-1}$. The effective
exposure time was 136.5~s and 121.3~s for the May and August observation,
respectively.}
 
\end{figure}

\newpage

\begin{figure}[H]
\centerline{\includegraphics[width=20cm]{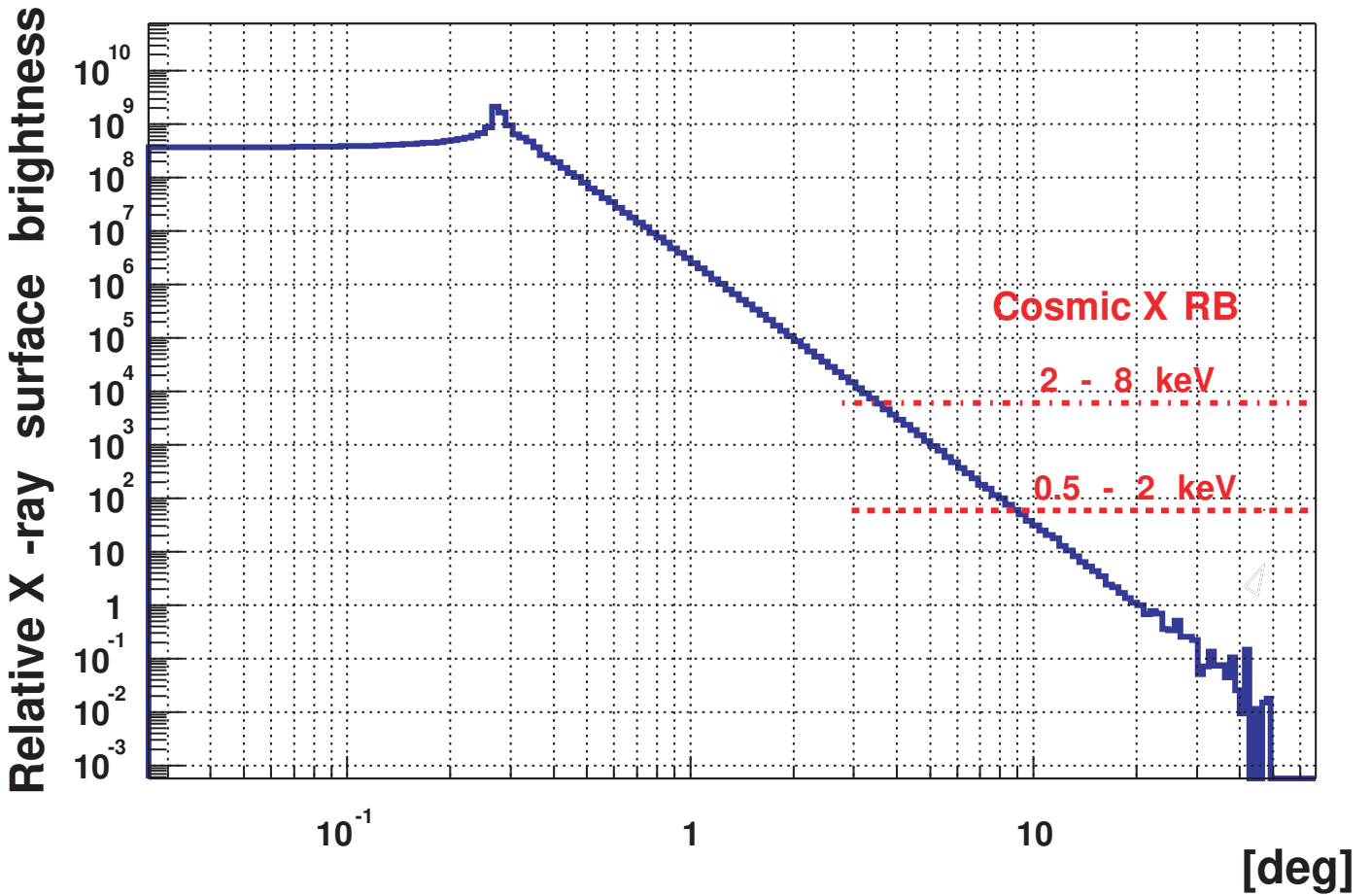}}
\caption[]{The predicted solar X--ray brightness from the Sun direction. The
level of the cosmic X--ray background radiation in the soft and hard energy
band is also shown \citep{cowie,moretti} defining the size of the observable
X--ray halo due to decaying solar axions (see text).}
\end{figure}


\vskip1.0cm
\begin{figure}[H]
\caption[]{See\,
{\tt http://www.kluweronline.com/CD/SOLA/210/1-2/Kruckeretal/fig1.pdf}\\
The GOES and RHESSI (from top to bottom) X--ray observations during $\sim 1$
hour of low solar activity. The vertical  lines define the spacecraft
day. The RHESSI detector background level is measured before and after the
daylight part of the orbit (dashed line). The RHESSI count spectrogram plot (bottom) is background subtracted. 
This Figure is taken from \citet{rhessix3}.}
\end{figure}

\end{document}